\documentstyle[aps,twocolumn]{revtex}
\draft
\title{Photon-photon interactions in cavity electromagnetically induced
transparency}
\author{M.~J. Werner and A. Imamo\=glu}
\address{Department of Electrical and Computer Engineering and Physics,
University of California, Santa Barbara, CA 93106-9560.}
\date{\today}

\begin{document}

\maketitle
\begin{abstract}
Dissipation-free photon-photon interaction at the single photon
level is studied in the context of cavity electromagnetically induced
transparency (EIT).
For a single multilevel atom exhibiting EIT in the strong cavity-coupling
regime,
the anharmonicity of the atom-cavity system has an upper bound determined
by single
atom-photon coupling strength. Photon blockade is inferred to occur for
both single
and multi-atom cases from the behaviour of transition rates
between dressed states of the system. Numerical calculations of the second
order
coherence function indicate that photon antibunching in both single and
two-atom
cases are strong and comparable.

\end{abstract}

\narrowtext

Enhancement of dissipation-free photon-photon interactions at the few photon
level is a fundamental challenge in quantum optics.
Such interactions are necessary for single photon quantum control as
in quantum logic gates. Typically, tuning close to
atomic resonances simultaneously increases both (Kerr) nonlinearity and linear
absorption, severely limiting the available nonlinear phase-shift.
In Ref.~\cite{kerr1}, a new method that is based on electromagnetically
induced transparency (EIT) \cite{EIT} was proposed.
In this scheme linear susceptibility vanishes on resonance, making
it possible to obtain as much as 8
orders-of-magnitude improvement in Kerr nonlinearity as compared to
conventional schemes based on three-level atoms.
Recently, experimental realization of this
scheme with an enhancement approaching $10^7$ using a
sodium Bose-Einstein-condensate (BEC) has been reported \cite{Hau}. In light
of these developments, it is natural to explore the limits on achievable
photon-photon interaction while maintaining EIT.

A single photon in a cavity can block the injection of a second
photon due to a photon blockade effect~\cite{photon-prl,Walls}. The
original proposal for using a cavity-EIT medium to achieve
photon blockade was based on an adiabatic elimination of the atomic degrees
of freedom~\cite{photon-prl}. Using a linearized analysis, Grangier et.
al.~\cite{grangier} have shown that the adiabatic elimination carried out in
Ref.~\cite{photon-prl} is not justified due to large dispersion of the EIT
medium. The correct description of photon-photon interactions in the single
photon limit requires a nonlinear analysis that treats the atom-photon
interactions exactly: this is the principal goal of  this Rapid Communication
where the analysis is based on the exact energy eigenstates of the coupled
atom-cavity system.
Under conditions of strong coupling, the anharmonicity in the spectrum of the
atom-cavity molecule can lead to nonclassical photon statistics
such as photon antibunching or a photon blockade effect.
The latter is determined by the probability of two photons absorbed
sequentially into the system while the former is related to the conditional
probability of photon emission given a single photon
was already emitted.
Both effects strongly depend on the anharmonicity of the atom-cavity energy
eigenstates in the doubly excited ($n=2$) manifold, or equivalently, how well
can the coupled system, under EIT conditions, be described as an effective
two-level system~\cite{cavityqed}.
Even though our analysis focuses on determining the level
shifts and linewidths in the $n=2$ manifold, we complement this analysis
with a calculation
of the second-order coherence function of the transmitted field.

We investigate the dependence of the photon-photon interaction on the atom
number, dispersion, and the atom-cavity coupling strength. We concentrate on
the situation where the cavity contains at most two photons, the atom-cavity
system exhibits EIT at  the cavity mode frequency, and a weak external driving
field is resonant with the EIT transition so that the only state excited  by
the first photon (i.e. in the $n=1$ manifold) is the cavity-EIT state.
The crucial advantages of this EIT scheme are absence of one photon
loss due to spontaneous emission, arbitrarily reduced cavity decay rate for
the $n=1$ manifold cavity-EIT state, and a multiatom induced photon-photon
interaction which, under
conditions of non-resonant EIT, can be comparable to a single atom.
In the context of quantum logic gates, atom-atom
interactions  between two multilevel atoms induced via a single cavity mode
have been discussed previously~\cite{vanEnk} where the role of photons and
atoms is reversed compared to our scheme.

The definition of photon blockade suggests
that its strength can be quantified by the transition rate $W_{1\rightarrow
2}$ from
$n=1$ manifold to $n=2$ manifold, normalized to the $n=0\rightarrow n=1$
transition rate. Assuming a single relatively stable state in the
$n=1$ manifold, we can measure the strength of the photon blockade effect by
$P=\sum_iW_{1\rightarrow 2i}/W_{0\rightarrow 1}$, which in turn suggests that
one way to analyze the problem is to calculate transition rates between exact
dressed-states. An experimentally more relevant quantity is the
degree of photon antibunching determined by
$g^2(0)=\langle{\hat a}^{\dagger^2} {\hat a}^2\rangle/
\langle{\hat a}^\dagger \hat a\rangle^2$. Even though these quantities are
physically different as described earlier, when the coupled system behaves like
a single two-level system $g^2(0)\ll 1$ implies photon blockade ($P\ll 1$).

Consider the $n=1$ manifold of the coupled atom-cavity system
where the cavity mode contains at most one photon. Since we assume throughout
this paper that the EIT condition is satisfied, the frequencies of the
classical coupling field ($\omega_c$) and the cavity mode ($\omega_{cav}$)
are chosen to satisfy two-photon resonance
$\omega_{cav} - \omega_c - \omega_{21} = \omega_{31}-\delta-
\omega_c -\omega_{21} = 0$ (see Fig.~1). The EIT condition
implies that there is no population  in the upper state
$|3\rangle$ which is the only atomic state with a significant
decay rate relevant to the $n=1$ manifold.
The effective interaction-picture Hamiltonian which couples $N$
4-level atoms with the single cavity mode
and the  coupling field with Rabi frequency $2\Omega$ is
\FL\begin{eqnarray}
\hat H/\hbar&=& -i\kappa \hat a^\dagger \hat a
+\sum_{i=1}^N\biggl[-i{\widetilde\Gamma_4}\hat\sigma_{44}^i
-i{\widetilde\Gamma_3}\hat\sigma_{33}^i + \Omega(\hat\sigma_{32}^i+
\hat\sigma_{23}^i)\nonumber\\ &+&g_{13}(\hat a \hat \sigma_{31}^i
+\hat a^\dagger\hat\sigma_{13}^i)
+g_{24}(\hat a \hat\sigma_{42}^i+
\hat a^\dagger \hat\sigma_{24}^i) \biggr] \ ,
\end{eqnarray}
where $\gamma_{31},\gamma_{32},\gamma_4$ are atomic level spontaneous
emission rates,
$\widetilde\Gamma_3=(\gamma_{31}+\gamma_{32})/2+i\delta$,
$\widetilde\Gamma_4=\gamma_4/2+i\Delta$, $\kappa$ is the cavity decay rate and
$\Delta = \omega_{41}-\omega_{21} - \omega_{cav}$.
The externally measured photon statistics of the driven cavity are
determined by the master equation
$\dot{\rho} = -i[\hat H/\hbar,\rho] + {\cal E}_p\bigl[\hat a^{\dagger}-\hat
a,\rho \bigr]
+ \hat{\mathcal J}\rho$
where $\hat{\mathcal J}$ is the jump superoperator for Lindblad form and
${\cal E}_p$ is the strength of the weak external drive field resonant with the
cavity mode.
The last term of Eqn.1 can be replaced by
$H_{perturb}=-\hbar\chi{\hat a}^\dagger {\hat a} \sum_i{\hat\sigma}_{22}^i $
in the limit $|\Delta| \gg g_{24},\Gamma_4$ where $\chi=g_{24}^2/\Delta$.

In the case of a single atom in the cavity ($N=1$), the $n=1$ and $n=2$
manifolds (in
the perturbative limit) have 3 eigenstates.
In the limit $\widetilde\Gamma_3=0$, eigenvalues
$\hbar(\omega_{cav}+\epsilon_i)$
for the $n=1$ manifold are approximately given by
\begin{equation}
\epsilon^{n=1}_{0}\approx\frac{-i\kappa}{1+\alpha} \ ,
\epsilon^{n=1}_{\pm}\approx-i\frac{\kappa}{2}\pm\Omega\sqrt{1+\alpha -
(\kappa/2\Omega)^2}
\end{equation}
where $\alpha=g_{13}^2 N/\Omega^2$ and $Re(\epsilon^{n=1}_{0})$ is
independent of $(g_{13},\Omega)$ and is on cavity resonance. The eigenstate
corresponding
to $\epsilon^{n=1}_{0}$
\begin{equation}
|\phi^{n=1}_0\rangle=
{1\over{\sqrt{1+\alpha}}}
\biggl[\hat a^\dagger-\frac{g_{13}}{\Omega}\hat\sigma_{21}\biggr]|0\rangle
\end{equation}
contains no component from
the upper atomic state $|3\rangle$ and in the limit
$\alpha \gg 1$ is predominantly atomic in character: this is
the cavity-EIT trapping state. The transition rate induced by a weak probe
from the ground state into $\epsilon^{n=1}_{0}$ is independent of $\alpha$.
In the $n=2$ manifold, in the limit $g_{24}=0$, one of the eigenstates is
\begin{equation}
|\phi^{n=2}_0\rangle={1\over{\sqrt{2+\alpha}}}
\hat a^\dagger\biggl[\hat
a^\dagger-\frac{g_{13}}{\Omega}\hat\sigma_{21}\biggr]|0\rangle \ .
\end{equation}
whose linewidth due to cavity decay has a weak $\alpha$ dependence of
$\kappa(4+\alpha)/(2+\alpha)$.
For $g_{24} \neq 0$, this
energy eigenstate experiences an ac-Stark shift.
Once the EIT supporting state $|\phi^{n=1}_0\rangle$ is excited, the
state closest to two-photon resonance for
$\delta = 0$ ($|\phi^{n=2}_0\rangle$) has a perturbative level shift due to
$g_{24}$ of
\begin{equation}
\epsilon^{n=2}\approx - \frac{g_{24}^2}{\Delta}
\frac{2g_{13}^2}{2g_{13}^2+\Omega^2}=
-\chi \frac{2\alpha}{1+2\alpha} \ , (N=1)
\end{equation}
allowing photon-blockade and photon antibunching to occur, provided $ \chi >
\kappa$. Level shift is largest in the limit $\alpha \gg 1$, which can be
achieved by reducing $\Omega$. One should note that additional constraints  on
$\Omega$ exist to  maintain EIT but are weak for cold alkali atoms~\cite{Hau}.

The results outlined above only consider the regime
where level $|4\rangle$ can be adiabatically eliminated. In this perturbative
limit, we found that  the
maximum value of the photon-photon interaction (as measured by the
anharmonicity or level shift in analogy with an ideal nonlinear cavity)  is
independent of $\alpha$ and is given by
$|\chi_{max,pert}|\sim |\chi|$.  This result suggests that the upper bound for
the non-perturbative photon-photon  interaction will be given by
$g_{24}$ in the single-atom case. Consider energies
$\hbar(2\omega_{cav}+\epsilon)$ of the $n=2$ manifold eigenstates.
In a nonperturbative single atom regime, there is no evidence that detunings
increase the anharmonicity.
The limiting case of vanishing detunings and losses gives:
\begin{equation}
\epsilon=
\pm\sqrt{\frac{G^2}{2}\pm
\sqrt{\frac{G^4}{4}-2g_{13}^2g_{24}^2}}
\end{equation}
where $G^2=g_{24}^2+\Omega^2+2g_{13}^2$.
For the resonant case ($\Delta=\delta=0$) with $g=g_{13}=g_{24}$,
the eigenvectors can be found exactly and eigenstates whose energy level is
closest to
$2\hbar\omega_{cav}$ have level shifts which
asymptotically  approach $g$ from below as $g$ increases. We believe that
this result
strongly suggests that the maximum possible anharmonicity
achievable in
cavity-EIT system is always bounded above by the single atom-cavity
coupling coefficient $g_{24}$.
When $g_{24}\sim \sqrt{2}g_{13}$ the non-perturbative nature shows itself
through level anti-crossings.
There is then a second upper bound given by $\sqrt{2}g_{13}$.

In order to relate these ideas to well known photon correlation measurements
we calculate $g^2(\tau)$ for experimentally achievable single atom parameters
(Fig.~2) \cite{experiments}. For weak cw driving such that the average
cavity  photon number is much less than unity, we expect $g^2(0)\ll1$ for
sufficiently large energy shifts in the $n=2$ manifold. If the coupled system
under weak excitation behaves ideally like a single two-level system then no
amplitude for photon modes with $n>1$ will contribute to
$g^2(0)$. One can understand this simply as restating the fact that photon
blockade implies that ideally, photons transmitted out of the cavity should
only  correspond to cavity decay of $|\phi_0^{n=1}\rangle$.
Components from other transitions can be strongly
suppressed  by large splittings of the dressed states.
The main contribution for nonzero $g^2(0)$ would then come from
$|\phi_0^{n=2}\rangle$.
However, the values
we find for $g^2(0)$ are as small as $2\mbox{x}10^{-3}$ for parameters from
recent experiments~\cite{caltech} on cavity-QED
($g=7.5\kappa,\gamma_{31}=0.325\kappa$).
Such experiments
might be enhanced using EIT. For $\Omega=2.5\gamma_{31}$ so that $\alpha\sim
10^2$ one would expect a reduction of the cavity-EIT decay rate
from their measured cavity decay rate of 16 MHz to 0.19 MHz.
Furthermore, future improvements in cavity and atom trap design
will lead to cavity-QED
experiments limited by the atomic spontaneous emission rate; cavity-EIT
schemes will be especially favourable in such a case.
Clearly the ultra-low values of $g^2(0)$ can not be explained by the level
shifts alone.
Since the weak probe is tuned exactly between the two dressed states in the
$n=2$ manifold
quantum interference might be responsible.

We next discuss how in a high dispersion regime
the usual linear scaling of the nonlinearity with the number of atoms does
not apply.
The $n=1$ manifold, where a single photon excitation is shared among the
$N$ atoms,
contains three energy levels
$\hbar(\omega_{cav}+\epsilon_i)$ whose energy shifts and widths, neglecting
spontaneous emission, are approximately given by Eqn.(2) provided we assume
that all atoms couple
identically to the cavity mode. Note that the eigenvalue $\epsilon^{n=1}_{0}$
has a width which decreases with the number of atoms
$N$ \cite{Lukin} while the splitting of the other two dressed states increases
with $N$. The eigenstate corresponding to $\epsilon^{n=1}_{0}$ is
\begin{equation}
|\phi^{n=1}_0\rangle=
{1\over{\sqrt{1+\alpha}}}
\biggl[\hat a^\dagger-\frac{g_{13}}{\Omega}
\sum_{i=1}^N\hat\sigma^i_{21}\biggr]|0\rangle \ .
\end{equation}
When $N>1$, the $n=2$ manifold contains six energy eigenstates (in the
perturbative
limit). For $g_{24}=0$, one of these eigenstates
\begin{equation}
|\phi^{n=2}\rangle=\frac{1}{\sqrt{2}(1+\alpha)}
\bigg[ {\hat a}^{\dagger} - \frac{g_{13}}{\Omega}\sum_{i=1}^N
\hat\sigma_{21}^i\bigg]^2 |0\rangle \ .
\end{equation}
has energy $\hbar 2 \omega_{cav}$, implying that the energy level diagram, as
probed by the weak external field, is harmonic.
The state $|\phi^{n=2}\rangle$ corresponds to two cavity-EIT trapping state
excitations
and has a linewidth which decreases with $1+\alpha$. For $g_{24} \neq 0$, this
energy eigenstate experiences an ac-Stark shift. When there are no degenerate
eigenstates the level shift of $|\phi^{n=2}\rangle$ caused by the perturbation
$H_{perturb}$ is
\begin{equation}
\epsilon^{n=2}\approx -2 \chi\frac{\alpha}{(1+\alpha)^2} \ .
\end{equation}
The atom-cavity molecule therefore has an anharmonic response to the drive
field, provided that $|\epsilon^{n=2}| > 2 \kappa/(1+\alpha)$. We note that the
(nonlinear) splitting of the eigenstate corresponding to
$\epsilon^{n=2}$ initially increases with the number of atoms $N$ in the
cavity as would be expected from a traditional nonlinear optical
system. However, as dispersion becomes important ($\alpha \simeq 1$),  this
increase saturates.  In the  highly dispersive limit given by $\alpha\gg 1$,
the splitting of the eigenenergy $\epsilon^{n=2}$ and its width decreases
with increasing N, under the assumption that all $N$ atoms have the maximum
possible  interaction.

As we have already argued, the photon-photon interaction strength
 is determined by
the energy level splittings and linewidths in the $n=2$ manifold, as well
as the
transition matrix elements. In the special case of
doubly resonant EIT ($\delta=0$) with $N>1$,  one of the eigenstates
\begin{equation}
|\psi^{n=2}\rangle = \frac{1}{\sqrt{6}}\bigg[ {\hat a}^{\dagger^2}
+\frac{1}{N}\sum_{i\neq j}\hat\sigma_{21}^i\hat\sigma_{21}^j
+\frac{1}{N}\sum_{i\neq j}\hat\sigma_{31}^i\hat\sigma_{31}^j\bigg] |0\rangle
\end{equation}
has energy $\hbar 2\omega_{cav}$ and a linewidth that is independent of
$\alpha$. This state  remains unshifted by
$g_{24}$  and has a nonzero transition rate from
$|\phi^{n=1}\rangle$ that decreases with $\alpha$. Degenerate perturbation
theory shows that the actual shifted eigenstate is then a linear combination of
$|\psi^{n=2}\rangle$ and
$|\phi^{n=2}\rangle$  with level shift $3\epsilon^{n=2}/2$.
The transition rate from the $n=1$ manifold EIT state to this shifted
eigenstate  increases with the  number of atoms $N$,
for $\alpha > 1,\chi>\alpha\kappa$.
Increasing the number of atoms therefore
leads to a reduction in the magnitude of the photon-photon interaction. The
presence of these two states implies that photon blockade is  small for
this special case of $\delta=0$, where one would naively expect the
nonlinearities
to be the strongest.
It is also interesting to note that the detrimental effect of the
degenerate state $|\psi^{n=2}\rangle$ that inhibits nonlinearity implies that
the adiabatic elimination of atomic degrees of freedom is not justified even
in the low dispersion limit
where it is normally assumed to be valid.

When $|\epsilon^{n=2}|,\kappa<|\delta|<\sqrt{Ng_{13}^2+\Omega^2}$, the
degeneracy
in the $n=2$ manifold is removed
and the dominant contribution for transition from $n=1$
to $n=2$ manifolds is determined by $|\phi^{n=2}\rangle$. In this case,
$P$ becomes independent of
$\alpha$ for $\alpha\gg 1$ and approaches the single-atom result.
Photon blockade experiments using the cavity-EIT system with $N\gg 1$
can therefore achieve
a reduction in two-photon absorption by the atom+cavity molecule comparable
to that of a
single atom+cavity.  However, the multiatom case is considerably more sensitive
to other effects not included in the dressed-state calculations: for example,
effective Rabi frequency associated with the external drive has to be much
smaller than $\epsilon^{n=2}$,  which places a constaint on ${\cal E}_p$.
Also, the spatial dependence of the atoms, their finite
velocities or atom-atom interactions may become more important.
A recent analysis of the spatial dependence of atomic beams~\cite{carmichael99}
for the Jaynes-Cummings model(JCM) where energy levels are proportional
to $G_{JCM}=\sqrt{\sum_{i=1}^N (g_i/g_{max})^2}$,
shows the probability distribution $P(G_{JCM})$ is well approximated by a
Gaussian with mean $\sqrt{N_{eff}}$, given by the atomic density
and cavity parameters, when $N_{eff}\gg 1$.
In the limit $N_{eff}\gg 1$ and $\alpha \gg 1$, the variation
in $\epsilon^{n=2}$ level splitting and linewidth
over the width of the distribution $P(G_{JCM})$ both decrease like
$1/\alpha^2$.
The cases $N_{eff}\sim 1$ and
$\alpha \sim 1$ are more involved and will be reported elsewhere.

The structure of the eigenstates and energy eigenvalues
is the same for any $N>1$ and so we have calculated $g^2(0)$ for $N=2$ versus
detuning $\delta$ of the cavity mode from $\omega_{31}$
while maintaining EIT using the
coupling constants and decay rates as in the single atom case (see Fig.~3).
The energy level splittings in the $n=1$ manifold constrains the detuning
so that
$\delta < \sqrt{g_{13}^2N+\Omega^2}$; we find that large $\delta$ also
results in
suppression of  energy level splittings in the $n=2$ manifold.
As can be seen in Fig.~3 values for $g^2(0)$ using two atoms
varies by three orders-of-magnitude over a
detuning $O(g)$; we find that for
such detunings $g^2(0)\ll 1$ and its value is of the same order-of-magnitude as
for a single atom for the same driving strength.
In contrast, when $\delta=0$, there is practically no photon
antibunching for the two-atom case - a result we anticipated based on  the
appearance of
the unshifted eigenstate $|\psi^{n=2}\rangle$.
Furthermore, for $|\delta|=g$,
${\cal E}_p=0.1\kappa$, $\Omega$ in the range $2\kappa\sim10\kappa$ we find
$g^2(0)
\simeq 10^{-3}$. As long as the driving satisfies ${\cal E}_p<
\kappa/(1+\alpha)$ then large antibunching should be robust. These values
for $g^2(0)$
are about two-orders-of-magnitude smaller than for similarly prepared
two-level atoms in
a cavity.

In summary, we have shown that photon-photon interactions using cavity-EIT
is experimentally feasible, and has distinct advantages over two-level atoms.
We used the  eigenstates of the perturbed many 3-level atom system
to give the $N$ atom dependence of the energy level shift.
Using a heuristic argument based on transition rates between dressed states we
inferred that multiatom photon blockade occurs provided non-resonant EIT is
used.
We have shown explicitly that photon antibunching for a multiatom case (two
atoms)  can
be comparable to that a single atom. We also showed that in the single-atom
case,
anharmonicity is bounded above by the single atom-photon coupling
strength $g_{24}$. This bound is not reached if $\sqrt{2}g_{13}<g_{24}$.

This work is supported by a David and Lucile Packard
Fellowship. After completion of this work, we became
aware of the preprint for
\cite{Rebic1998}, which provides further analyses of the single atom
case  using a calculation of $g^{(2)}(\tau)$ as in [6].
A.I thanks Stojan Rebic for useful discussions.
Numerical work supported in part by National Science Foundation grant
CDA96-01954, and by Silicon Graphics Inc.

\begin{figure}
\caption{(a) Internal atomic degrees of freedom
of a four-level atom coupled to a single cavity mode with frequency
$\omega_{cav}$
and an external laser field characterized by the Rabi frequency $2\Omega$. (b)
The energy levels of the coupled atom-cavity system.}
\end{figure}
\begin{figure}
\caption{Second-order photon correlation $g^2(\tau)$
demonstrating photon anti-bunching.
The coupling constant and decay rates correspond to current single atom
experiments~[10] of $g_{13}=g_{24}=7.5\kappa$,
$\gamma_4=\gamma_{31}=\gamma_{32}=0.325\kappa$
 and $\delta=\Delta=0,\Omega=2.5\gamma_{31},{\mathcal E}_p=0.1\kappa$.}
\end{figure}
\begin{figure}
\caption{Two atom case ($N=2$) showing the variation of $g^2(0)$
with the detuning $\delta=\omega_{31}-\omega_{cav}$
for the cavity-EIT giant Kerr nonlinearity.
The coupling constant and decay
rates are the same as in Fig.~2 while the detunings and drive strengths are
$\Delta=0,{\mathcal E}_p=0.01\kappa$.}
\end{figure}

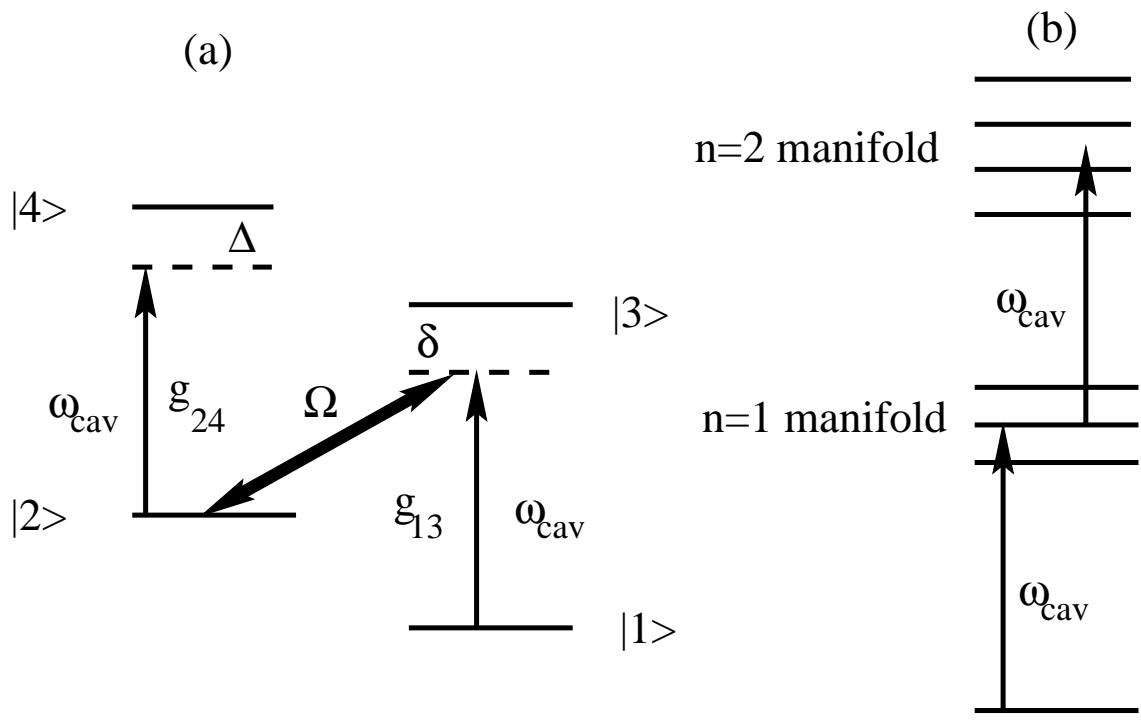
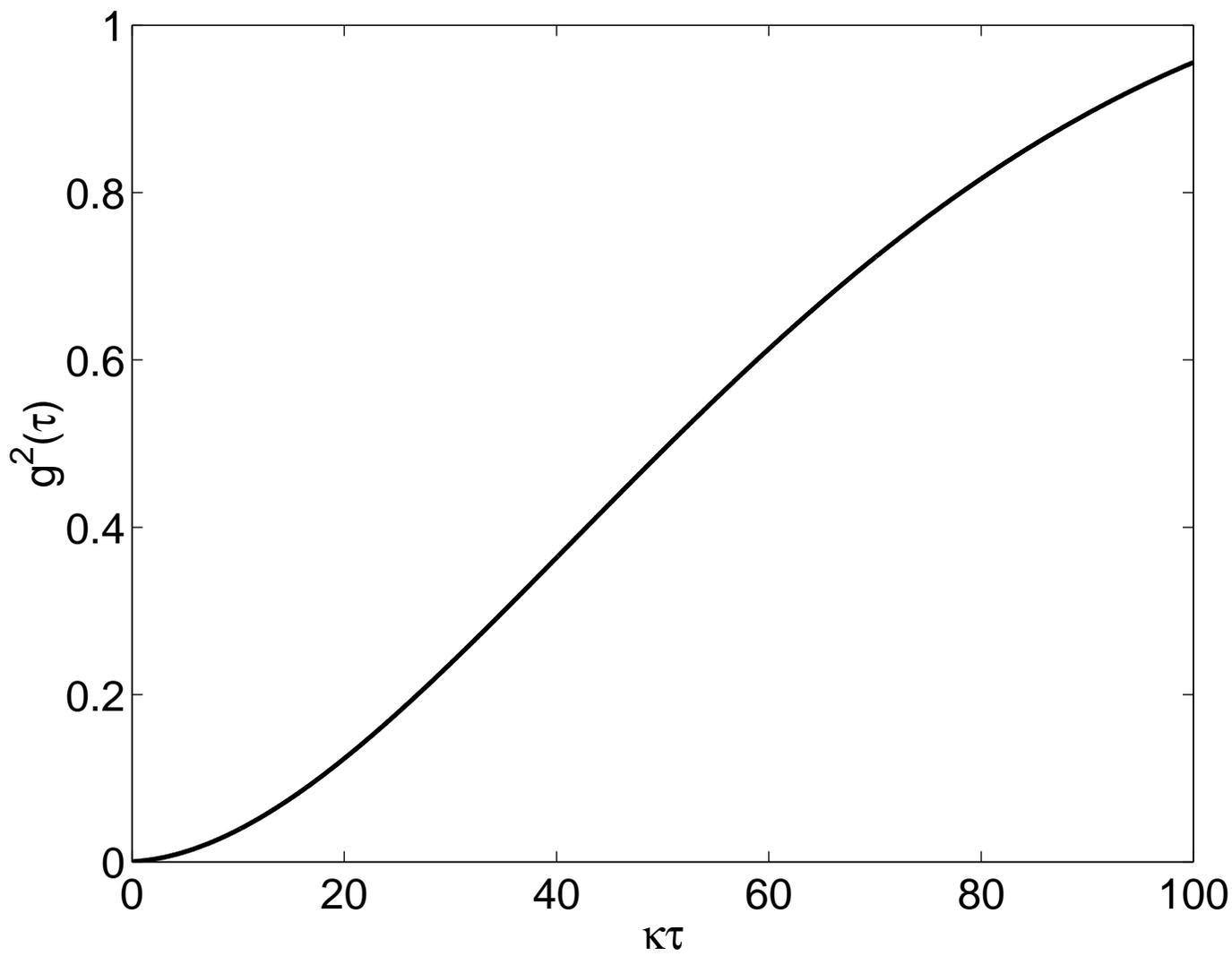
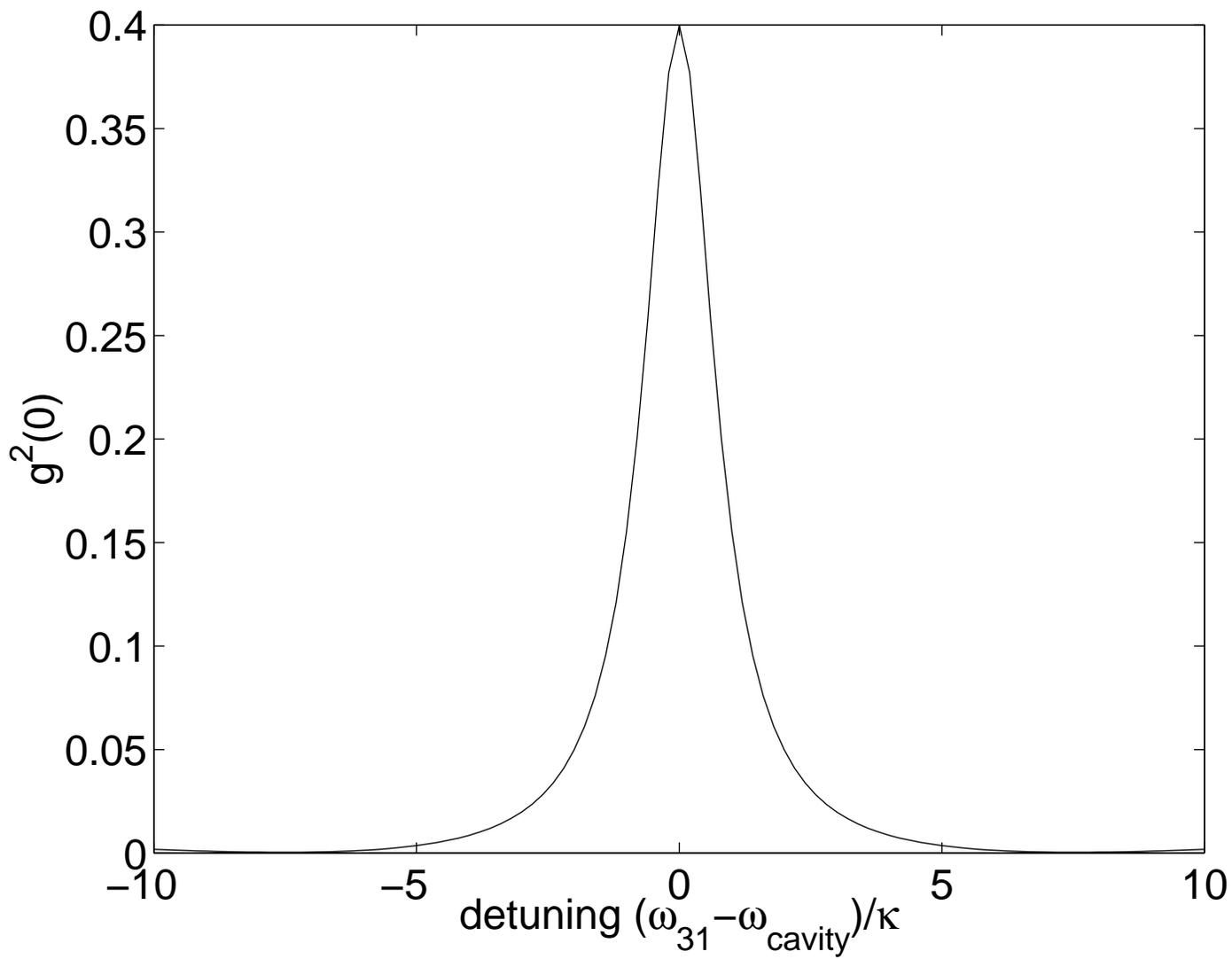
\end{document}